\documentclass[12pt,letterpaper]{article}

\usepackage[letterpaper,margin=1in]{geometry}

\usepackage{newtxtext,newtxmath}  

\usepackage{setspace}
\makeatletter
\renewcommand\footnotesize{\@setfontsize\footnotesize{10pt}{12pt}}
\makeatother

\usepackage{fancyhdr}
\usepackage{graphicx}
\usepackage{multirow}

\makeatletter
\@ifundefined{Bbbk}{}{}
\makeatother

\usepackage{amsmath,amssymb,amsfonts,mathtools}

\DeclareMathOperator{\Var}{Var}
\DeclareMathOperator{\Cov}{Cov}

\newcommand{\E}{\mathbb{E}}

\DeclareMathOperator{\doop}{do}   



\usepackage{amsthm}

\usepackage{mathrsfs}
\usepackage[title]{appendix}
\usepackage{xcolor}
\usepackage{textcomp}
\usepackage{enumitem}
\usepackage{booktabs}
\usepackage{algorithm}
\usepackage{algorithmicx}
\usepackage{algpseudocode}
\usepackage{comment}
\usepackage{listings}
\usepackage{tikz}
\usepackage{hyperref}
\usepackage{caption}
\usetikzlibrary{positioning,calc,arrows.meta}
\usetikzlibrary{arrows.meta,positioning,fit,shapes.geometric,calc}
\usepackage{subcaption}
\usepackage{adjustbox}   
\tikzset{dot/.style={circle,fill=black,inner sep=1.2pt}, >=Latex}
\usepackage[backend=biber,style=authoryear,maxcitenames=2,maxbibnames=6,doi=false,isbn=false,url=false]{biblatex}
\title{Rethinking Beta: A Causal Take on CAPM\thanks{Preprint. Under review.}}

\author{Naftali Cohen\thanks{The views expressed are solely the author’s and do not necessarily reflect those of any current or former employer or affiliated institution. This paper was prepared in a personal capacity, without the use of employer resources or confidential information. No investment advice is offered. Any errors are the author’s alone.}\\
Independent Researcher\\ New York, NY\\
\texttt{naftalic@gmail.com}}
\date{\today}
\date{}

\begin{document}
\begingroup
\renewcommand\thefootnote{\fnsymbol{footnote}} 
\maketitle
\endgroup



\abstract{
The CAPM regression is typically interpreted as if the market return contemporaneously \emph{causes} individual returns, motivating beta–neutral portfolios and factor attribution. 
For realized equity returns, however, this interpretation is inconsistent: a same–period arrow $R_{m,t}\!\to\!R_{i,t}$ conflicts with the fact that $R_m$ is itself a value–weighted aggregate of its constituents, unless $R_m$ is lagged or leave–one–out—the “aggregator contradiction.” 

We formalize CAPM as a structural causal model and analyze the admissible three–node graphs linking an external driver $Z$, the market $R_m$, and an asset $R_i$. 
The empirically plausible baseline is a \emph{fork}, $Z\!\to\!\{R_m,R_i\}$, not $R_m\!\to\!R_i$. 
In this setting, OLS beta reflects not a causal transmission, but an attenuated proxy for how well $R_m$ captures the underlying driver $Z$. 
Consequently, “beta–neutral” portfolios can remain exposed to macro or sectoral shocks, and hedging on $R_m$ can import index-specific noise.

Using stylized models and large–cap U.S.\ equity data, we show that contemporaneous betas act like proxies rather than mechanisms; any genuine market–to–stock channel, if at all, appears only at a lag and with modest economic significance. 

The practical message is clear: CAPM should be read as associational. 
Risk management and attribution should shift from fixed factor menus to explicitly declared causal paths, with “alpha’’ reserved for what remains invariant once those causal paths are explicitly blocked.
}





\section{Introduction: Why Causality Matters for Portfolio Management}\label{sec:intro}

In day-to-day portfolio management, ``beta'' is more than a regression coefficient---it is treated as a control knob for market exposure.
Practitioners routinely run beta-neutral books precisely to keep broad market moves from dominating P\&L \parencite{chincarini2006qepm,zhou2014active,isichenko2021qpm,paleologo2021advanced,paleologo2025elements}. This operational view is typically instantiated through the CAPM regression\footnote{Our analysis applies to both the time-series and pooled (panel) regression views of CAPM.},
\[
R_i \;=\; \alpha \;+\; \beta\,R_m \;+\; \epsilon_i,
\]
whose intellectual roots go back to \textcite{sharpe1964capm} and \textcite{lintner1965valuation}.
In practice, $\alpha$ and $\beta$ are estimated by OLS on rolling windows---often 36–60 months of monthly data or multi-year daily samples---following conventions in empirical asset pricing and portfolio construction \parencite{fama1992crosssection,fama1993factors,chincarini2006qepm}. The informal interpretation most people carry is causal: if $\beta{=}1.2$, then ``a $1\%$ market rally causes a $1.2\%$ move in the stock, all else equal.'' This interpretation is convenient---but for realized returns, it is generally misleading.

\paragraph{Association vs.\ causation.}
OLS reports an association,
\[
\beta_{\text{OLS}} \;=\; \frac{\Cov(R_m,R_i)}{\Var(R_m)},
\]
whereas a causal effect is a counterfactual object,
\[
\beta_{\text{causal}} \;=\; \frac{\partial}{\partial r}\, \E\big[R_i \,\big|\doop(R_m=r)\big],
\]
defined by an intervention on $R_m$ \parencite{pearl2009causality,pearl2016primer,hernanrobins2020,de2023causal}. These two objects coincide only under specific graphical conditions: no back-door paths from $R_m$ to $R_i$, no simultaneity or feedback loops, correct time indexing, and a well-measured treatment. For the contemporaneous CAPM applied to realized equity returns, these conditions typically do not hold.

\paragraph{The hidden simultaneity in same-period CAPM.}
The market return is a value–weighted aggregate of its constituents; for an index such as the S\&P 500,
\[
R_{m,t} \;=\; \sum_{j} w_{j,t-1}\, R_{j,t}.
\]
A regression of a constituent on $R_{m,t}$ at the same time index therefore posits a two–equation system: the constituent helps determine the index and, purportedly, the index determines the constituent. This is exactly the simultaneity problem emphasized by \textcite{haavelmo1943simultaneous}, developed into a structural–probabilistic program by the Cowles Commission \parencite{koopmans1950statistical}, and addressed with system methods and instruments \parencite{theil1953repeated,sargan1958estimation}. \textcite{sims1980macroeconomics} later criticized ad hoc exclusions and advocated reduced–form, time–ordered VARs as a way to let the data speak before imposing structure. 
In finance, similar concerns have appeared in debates on identification and testability of asset pricing models \parencite{roll1977critique,shanken1992betapricing,stambaugh1997analyzing}.

\paragraph{Related literature.}
A large body of work has extended or critiqued CAPM by introducing time variation, conditioning information, or alternative mechanisms. Conditional CAPM models \parencite{jagannathan1996conditional,lewellen2006conditional} and dynamic conditional beta models \parencite{engle2016dynamic} explicitly allow betas to vary with state variables, while multi-factor and intertemporal extensions \parencite{merton1973intertemporal,fama1993factors} recognize that the market factor is incomplete. Other strands emphasize structural frictions such as leverage, funding constraints, and margin requirements as transmission channels \parencite{frazzini2014bab}. These approaches establish that betas can be state-dependent, incomplete, or shaped by frictions, and they remain central tools in empirical asset pricing. Yet none of them resolve the logical simultaneity created by defining the market as a contemporaneous aggregate of its constituents. Our contribution is complementary: by casting CAPM as an explicit structural causal model, we show that a same-period $R_m\!\to\!R_i$ arrow is incoherent unless lagged or leave-one-out, a perspective distinct from but consistent with earlier critiques of the market proxy problem \parencite{roll1977critique} and recent calls for explicitly causal factor models \parencite{de2023causal}.

\paragraph{Our contribution.}
This paper advances the CAPM debate on several fronts. 
First, we cast CAPM explicitly as a structural causal model using a minimal three-node core $(Z,R_m,R_i)$, making visible the assumptions that are only implicit in standard treatments. 
Second, we formalize what we call the \emph{Aggregator Contradiction}: because $R_{m,t}$ is defined as a value-weighted aggregate of its constituents, a same-period arrow $R_{m,t}\to R_{i,t}$ cannot co-exist with acyclicity unless the constituent is excluded (leave-one-out) or the causal effect is zero. This observation shows that the canonical CAPM regression cannot be read as a same-day structural equation. 
Third, we extend the analysis beyond this impossibility result by systematically examining the admissible three-node DAGs, articulating the economic mechanisms they correspond to, and assessing their plausibility in practice. The necessary-condition tests (temporal priority, acyclicity, mechanism, elimination) rule out some graphs while highlighting the fork $Z_t\to \{R_{m,t+\Delta},R_{i,t+\Delta}\}$ as the empirically dominant baseline.
Fourth, we demonstrate both analytically and empirically that in such forks the OLS $\beta$ is not a market-to-stock transmission coefficient but an attenuated proxy ratio, with residual exposure to $Z$ even after “beta-neutral” hedging. Example Gaussian SEMs make this algebraically transparent, and large-cap U.S. data confirm the patterns: slopes collapse once observable shocks are added, vary systematically by environment, and residuals continue to load on macro/sectoral forces. Finally, we translate these results into operational diagnostics—lead–lag tests, leave-one-out checks, shock-day attenuation—that practitioners can use to assess whether a given DAG is plausible for their portfolio.

\paragraph{Horizon and sampling.}
Intraday shocks often propagate within minutes to hours, so their effects can be fully reflected in end-of-day prices even when the actuation occurred at finer resolution. Our analysis intentionally avoids intraday sampling to rely on public data and to match widespread systematic practice. A higher-frequency replication (e.g., 5–30 minute bars with admissible lagging or leave-one-out construction) is a natural follow-up and would sharpen the timing diagnostics without changing the logical results derived here.

\paragraph{Why this matters for hedging and for “alpha’’.}
If $\beta$ is not causal, a “beta-neutral’’ portfolio can remain exposed to the underlying macro or sector driver $Z$ that moves both the market and the stock. The portfolio may look market-flat yet still bleed when $Z$ shifts; worse, mechanically hedging on $R_m$ can import index-specific noise into residual P\&L without removing the mechanism. Recent work has sharpened this critique: López de Prado argues that most factor models are associational, rarely declare a causal graph, and thus invite spurious claims and non-invariant alphas. He calls for explicit mechanisms, falsifiable implications, and simulated or natural interventions when experiments are infeasible \parencite{de2023causal}. This causal stance reframes factor neutrality and “alpha’’ as claims that must be justified by a graph, not simply taken from a factor menu.

\paragraph{A minimal causal lens.}
We therefore adopt a Minimal Causal Graph (MCG) with three nodes $(R_m,R_i,Z)$ and ask which directed acyclic configurations survive basic \emph{necessary conditions} (temporal priority, acyclicity, a concrete transmission mechanism, and a coherent ``causal elimination'' property) while respecting the fact that $R_{m,t}$ is an aggregate. This framing, grounded in structural causal models and invariance logic \parencite{pearl2009causality,pearl2016primer,peters2017elements,hernanrobins2020}, yields three operational consequences. 

First, the empirically plausible baseline is often a fork, $Z\!\to\!\{R_m,R_i\}$: in that case, OLS $\beta$ is a proxy exposure to $Z$, attenuated by how noisily $R_m$ measures $Z$, not a causal $R_m\!\to\!R_i$ coefficient. 

Second, when a market$\to$stock mechanism exists at all, it must respect time ordering (e.g., $R_{m,t-1}\!\to\!R_{i,t}$) or a leave–one–out construction ($R_{m,t}^{-i}$), and any such claim must be supported by lead–lag evidence rather than assumed. 

Third, hedging and “alpha’’ become DAG-dependent objects: the invariance you expect from a hedge only arises when you remove causal paths and block back-doors, not when you simply subtract a fixed list of factors.

In what follows, we make the implied structural model explicit, derive the population meaning of OLS under each admissible DAG, formalize the Aggregator Contradiction, and present numerical and empirical checks aligned with this causal perspective.


\section{CAPM as a Structural Causal Model}\label{sec:capm-scm}

To interpret CAPM causally one must first make its structural content explicit. In particular, we need an acyclic structural causal model (SCM) that specifies which variables act on which others, and under what timing and exogeneity restrictions. Only in such a model can a regression slope be interpreted as a path coefficient rather than a mere projection \parencite{pearl2016primer}.

\subsection{Assumed CAPM SCM}\label{subsec:capm-scm-assumed}

\begin{figure}[t]
\centering
\begin{tikzpicture}[>=stealth, node distance=1cm]
\node (Uz) {$U_Z$};
\node[below=of Uz] (Ux) {$U_X$};
\node[right=of Uz] (Z) {$Z$};
\node[right=of Ux] (X) {$X$};
\node[below right=.2cm and .8cm of Z](Y) {$Y$};
\node[above=of Y] (Uy) {$U_Y$};
\draw[dashed,->] (Uz) -- (Z);
\draw[dashed,->] (Ux) -- (X);
\draw[->] (Z) -- (Y);
\draw[->] (X) -- (Y);
\draw[dashed,->] (Uy) -- (Y);
\end{tikzpicture}
\caption{Canonical SCM for a causal reading of CAPM: $X\!\perp\! Z$, $X\!\perp\! U_Y$, and $Y$ has parents $\{X,Z\}$. Dashed arrows denote exogenous disturbances. Here $X\equiv R_m$ (market return), $Y\equiv R_i$ (asset $i$), and $Z$ collects non-market determinants of $Y$ (macro and/or sector/firm drivers).}
\label{fig:capm-scm}
\end{figure}
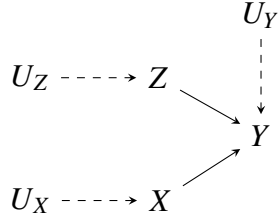

Let $X\equiv R_m$ (market return), $Y\equiv R_i$ (asset $i$), and let $Z$ collect non-market determinants of $Y$ (macro and/or sector/firm drivers). Over a single sampling interval $\Delta$ (time indices suppressed), introduce mutually independent exogenous shocks $(U_X,U_Y,U_Z)$ with $\E(U_\cdot)=0$, and posit the SCM
\begin{equation}
\label{eq:scm-capm}
Z := f_Z(U_Z),\qquad
X := f_X(U_X),\qquad
Y := f_Y\!\big(X,Z,U_Y\big),
\end{equation}
so that $X\!\perp\! Z$ and $X\!\perp\! U_Y$ (Fig.~\ref{fig:capm-scm}). A convenient linearization is
\begin{equation}
\label{eq:lin-capm}
Y \;=\; \beta_X\,X \;+\; \beta_Z\,Z \;+\; U_Y,\qquad \E(Z)=\E(U_Y)=0.
\end{equation}

Under \eqref{eq:scm-capm}–\eqref{eq:lin-capm}, back-door paths from $X$ to $Y$ are blocked and the causal and associational regressions agree:
\begin{equation}
\label{eq:do-equals-cond}
\E\!\left[Y \,\middle|\doop(X=x)\right] \;=\; \E\!\left[Y \,\middle|X=x\right] \;=\; \beta_X x,
\end{equation}
so the population OLS slope coincides with the direct causal effect,
\begin{equation}
\label{eq:ols-equals-betaX}
\beta^{\mathrm{OLS}}_{Y\sim X} \;=\; \frac{\Cov(X,Y)}{\Var(X)} \;=\; \beta_X,
\end{equation}
exactly in line with the back-door criterion \parencite[Ch.~3]{pearl2016primer}. These equalities require the absence of arrows $Z\!\to\!X$ and $Y\!\to\!X$, and the correct time ordering of $X$ relative to $Y$.

\paragraph{Model vs.\ measurement.}
The SCM in \eqref{eq:scm-capm}–\eqref{eq:lin-capm} is a coherent causal story only if the $X$ fed to the regression is admissible for the sampling scheme—for example, a lagged market return $R_{m,t-1}$ or a leave-one-out index $R_{m,t}^{-i}$. With the empirical same-time index typically used in practice, the story collides with acyclicity (simultaneity), a classic econometric concern \parencite{haavelmo1943simultaneous}; see \S\ref{subsec:assess}.

\subsection{Necessary conditions (testable implications)}\label{subsec:necessary1}
Any proposed edge $X\!\to\!Y$ must satisfy basic domain-testable implications \parencite{pearl2009causality,pearl2016primer,de2023causal}:

\emph{Temporal priority}: Changes in $X$ must precede and be able to influence changes in $Y$ at the operative horizon $\Delta$.

\emph{Acyclicity}: There is no same-period feedback $Y\!\to\!X$ that would render the system simultaneous at the measurement index.

\emph{Mechanism}: One must articulate a coherent economic transmission path by which $X$ could act on $Y$ (as opposed to both responding to a common driver).

\emph{Causal elimination}: The intervention $\doop(X{=}x)$ must be well-posed and, if enacted, remove an $X$-induced component of $Y$.

These are weak but useful filters: they do not identify an edge from observational data, but they do rule out incoherent specifications before estimation.

\subsection{Assessing CAPM's structure against realized-return measurement}\label{subsec:assess}
For realized returns the market is measured as a contemporaneous value-weighted aggregate,
\begin{equation}
\label{eq:aggregator}
X_t \;=\; \sum_{j} w_{j,t-1}\,Y_{j,t}.
\end{equation}
Read against the conditions in \S\ref{subsec:necessary1}, \eqref{eq:aggregator} clashes with a same-period arrow $X_t\!\to\!Y_{i,t}$ in four ways. 

First, temporal priority fails: $X_t$ is a deterministic function of $\{Y_{j,t}\}$ at the same time index.  

Second, acyclicity fails: adding $X_t\!\to\!Y_{i,t}$ closes a two-node cycle $Y_{i,t}\!\to\!X_t\!\to\!Y_{i,t}$, i.e., a simultaneous system.  

Third, the required mechanism is missing: economy-wide movers (policy surprises, risk appetite, liquidity shocks) act as common drivers $Z_{t-1}\!\to\!\{X_t,Y_{i,t}\}$, whereas the index is an aggregate statistic, not a physical actuator.  

Fourth, causal elimination is ill-posed: an intervention $\doop(X_t{=}x)$ would typically require simultaneous interventions on constituents—including $Y_{i,t}$—defeating isolation of an $X\!\to\!Y$ path.

\subsection{The Aggregator Contradiction (formal)}\label{subsec:agg-contr}
We now formalize the incompatibility between the empirical aggregator \eqref{eq:aggregator} and a same-period causal CAPM in an acyclic SCM.

\paragraph{Proposition (Aggregator Contradiction).}
Fix $t$. Suppose an acyclic SCM satisfies both (i) the aggregator identity
\[
X_t=\sum_j w_{j,t-1}Y_{j,t},\quad w_{j,t-1}\ge 0,\;\sum_j w_{j,t-1}=1,
\]
with $w_{i,t-1}>0$ and at least one $w_{j,t-1}>0$ for some $j\neq i$, and (ii) a same-period structural equation
\[
Y_{i,t}:=\beta\,X_t+Z_t,
\]
where $Z_t$ does not depend on $Y_{i,t}$ (it may depend on $\{Y_{j,t}:j\neq i\}$, exogenous shocks, and lags). Then acyclicity forces $\beta=0$. Equivalently, if $\beta\neq 0$ then either $w_{i,t-1}=0$ (leave-one-out index), all other weights vanish (a degenerate single-asset “market’’), or the system is simultaneous (cyclic) at $t$.

\begin{proof}
Substitute $X_t$ into $Y_{i,t}=\beta X_t+Z_t$ to obtain
\[
(1-\beta w_{i,t-1})Y_{i,t}
=\beta\sum_{j\neq i}w_{j,t-1}Y_{j,t}+Z_t.
\]
If $w_{i,t-1}>0$, at least one other $w_{j,t-1}>0$, and $\beta\neq 0$, then $Y_{i,t}$ depends contemporaneously on other endogenous variables $\{Y_{j,t}:j\neq i\}$, creating a simultaneous system at $t$ and violating acyclicity. This contradiction disappears only if $\beta=0$, if $w_{i,t-1}=0$, or if all other weights vanish.
\end{proof}

\paragraph{Corollaries.}
A causal reading of CAPM on realized returns is admissible only in three situations:
\begin{enumerate}
\item[(i)] \emph{Lagged market:} $X_{t-\Delta}\!\to\!Y_t$, which respects time ordering;
\item[(ii)] \emph{Leave-one-out index:} $X_t^{-i}=\sum_{j\neq i}w_{j,t-1}Y_{j,t}$, which removes $Y_{i,t}$ from the aggregator;
\item[(iii)] \emph{Degenerate case:} the “market’’ is just asset $i$ itself, i.e.\ $w_{i,t-1}=1$ and all other weights vanish.
\end{enumerate}
Cases (i) and (ii) yield coherent acyclic graphs but still require a concrete transmission mechanism consistent with timing and exogeneity (\S\ref{subsec:necessary1}). Case (iii) is trivial: the market reduces to a single asset, so the CAPM regression collapses to an identity and carries no substantive causal content.


\section{Possible causal CAPM DAGs}\label{sec:possible-dags}

If the contemporaneous graph in Fig.~\ref{fig:capm-scm} cannot represent how realized returns are measured, what---if anything---does an OLS regression of $R_i$ on $R_m$ recover? More broadly, which directed acyclic configurations over $(X,Y,Z)$ remain admissible once we respect time ordering and the fact that $R_{m,t}$ is an aggregate of constituents?

\paragraph{Graphs as hypotheses.}
We adopt a causal–modeling stance in which graphs are not data summaries but \emph{hypotheses} about data‐generating mechanisms. They encode domain assumptions and imply testable patterns of (in)dependence under conditioning and time shifts \parencite{pearl2009causality,pearl2016primer,peters2017elements}. OLS will always deliver a projection coefficient; whether that slope can be read as a causal path coefficient or merely as an associational proxy depends entirely on the underlying DAG and the time indices we feed it. Throughout, any interpretation of $\beta$ is therefore conditional on the stated graph and sampling scheme.

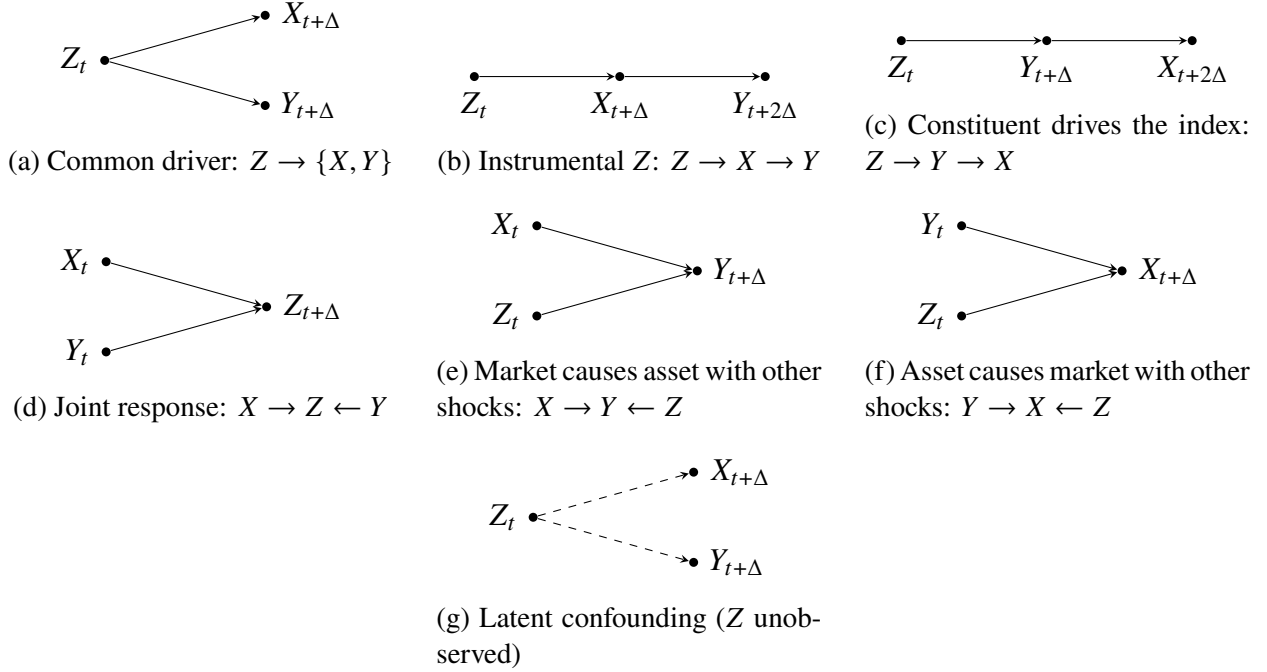
\begin{figure}[t]
\centering
\tikzset{dot/.style={circle,fill=black,inner sep=1.2pt}}
\begin{subfigure}{0.31\textwidth}
\centering
\begin{tikzpicture}[>=stealth, node distance=20mm]
\node[dot,label=left:{$Z_t$}] (Z) {};
\node[dot,label=right:{$X_{t+\Delta}$}, right=of Z, yshift=6mm] (X) {};
\node[dot,label=right:{$Y_{t+\Delta}$}, right=of Z, yshift=-6mm] (Y) {};
\draw[->] (Z) -- (X);
\draw[->] (Z) -- (Y);
\end{tikzpicture}
\caption{Common driver: $Z\to\{X,Y\}$}
\end{subfigure}
\hfill
\begin{subfigure}{0.31\textwidth}
\centering
\begin{tikzpicture}[>=stealth, node distance=18mm]
\node[dot,label=below:{$Z_t$}] (Z) {};
\node[dot,label=below:{$X_{t+\Delta}$}, right=of Z] (X) {};
\node[dot,label=below:{$Y_{t+2\Delta}$}, right=of X] (Y) {};
\draw[->] (Z) -- (X);
\draw[->] (X) -- (Y);
\end{tikzpicture}
\caption{Instrumental $Z$: $Z\to X\to Y$}
\end{subfigure}
\hfill
\begin{subfigure}{0.31\textwidth}
\centering
\begin{tikzpicture}[>=stealth, node distance=18mm]
\node[dot,label=below:{$Z_t$}] (Z) {};
\node[dot,label=below:{$Y_{t+\Delta}$}, right=of Z] (Y) {};
\node[dot,label=below:{$X_{t+2\Delta}$}, right=of Y] (X) {};
\draw[->] (Z) -- (Y);
\draw[->] (Y) -- (X);
\end{tikzpicture}
\caption{Constituent drives the index: $Z\to Y\to X$}
\end{subfigure}

\vspace{0.6em}

\begin{subfigure}{0.31\textwidth}
\centering
\begin{tikzpicture}[>=stealth, node distance=20mm]
\node[dot,label=right:{$Z_{t+\Delta}$}] (Z) {};
\node[dot,label=left:{$X_t$}, left=of Z, yshift=6mm] (X) {};
\node[dot,label=left:{$Y_t$}, left=of Z, yshift=-6mm] (Y) {};
\draw[->] (X) -- (Z);
\draw[->] (Y) -- (Z);
\end{tikzpicture}
\caption{Joint response: $X\to Z \leftarrow Y$}
\end{subfigure}
\hfill
\begin{subfigure}{0.31\textwidth}
\centering
\begin{tikzpicture}[>=stealth, node distance=20mm]
\node[dot,label=right:{$Y_{t+\Delta}$}] (Y) {};
\node[dot,label=left:{$X_t$}, left=of Y, yshift=6mm] (X) {};
\node[dot,label=left:{$Z_t$}, left=of Y, yshift=-6mm] (Z) {};
\draw[->] (X) -- (Y);
\draw[->] (Z) -- (Y);
\end{tikzpicture}
\caption{Market causes asset with other shocks: $X\to Y \leftarrow Z$}
\end{subfigure}
\hfill
\begin{subfigure}{0.31\textwidth}
\centering
\begin{tikzpicture}[>=stealth, node distance=20mm]
\node[dot,label=right:{$X_{t+\Delta}$}] (X) {};
\node[dot,label=left:{$Y_t$}, left=of X, yshift=6mm] (Y) {};
\node[dot,label=left:{$Z_t$}, left=of X, yshift=-6mm] (Z) {};
\draw[->] (Y) -- (X);
\draw[->] (Z) -- (X);
\end{tikzpicture}
\caption{Asset causes market with other shocks: $Y\to X \leftarrow Z$}
\end{subfigure}

\vspace{0.6em}

\begin{subfigure}{0.31\textwidth}
\centering
\begin{tikzpicture}[>=stealth, node distance=20mm]
\node[dot,label=left:{$Z_t$}] (Z) {};
\node[dot,label=right:{$X_{t+\Delta}$}, right=of Z, yshift=6mm] (X) {};
\node[dot,label=right:{$Y_{t+\Delta}$}, right=of Z, yshift=-6mm] (Y) {};
\draw[dashed,->] (Z) -- (X);
\draw[dashed,->] (Z) -- (Y);
\end{tikzpicture}
\caption{Latent confounding ($Z$ unobserved)}
\end{subfigure}

\caption{Seven admissible time–indexed DAGs for $(X,Y,Z)$. Arrows point left-to-right across time within each panel.}
\label{fig:seven-dags}
\end{figure}

\paragraph{Our approach.}
We propose a taxonomy of seven time–indexed DAGs involving a mechanistic driver $Z$ (Fig.~\ref{fig:seven-dags}). The list is illustrative rather than exhaustive: mechanisms may coexist, and the relevant graph can vary across horizons, assets, and market regimes. Our emphasis is on plausibility, not identification. For each DAG we (i) sketch the economic story it represents, (ii) state the population meaning of the OLS slope under that story, and (iii) note empirical diagnostics that could falsify it. Terms such as “likely’’ or “plausible’’ are to be read as context–dependent judgments informed by typical equity data and institutional constraints.

We use $Z$ to denote external drivers of returns, which may be firm–specific (e.g., earnings surprises, supply–chain shocks) or macro/market-wide (e.g., CPI releases, discount–rate or funding shocks). For compactness, let $\beta^{\mathrm{OLS}}_{k,\ell}$ denote the population slope in the regression of $Y_{t+k\Delta}$ on $X_{t+\ell\Delta}$ (plus an intercept), where $\Delta$ is the sampling interval.

\paragraph{Seven illustrative DAGs.}
\begin{enumerate}[label=(\alph*)]

\item \textbf{Fork (common driver): $Z_t \to X_{t+\Delta}$ and $Z_t \to Y_{t+\Delta}$.}  
Macro news, risk appetite, or liquidity shocks move many assets and---mechanically---the index. Large single-name shocks can also spill over via sector ETFs or correlated peers. Here $\beta^{\text{OLS}}_{1,0}$ reflects $Z$-induced co-movement, not a causal $X\!\to\!Y$ effect. Adding strong $Z$ proxies (macro-surprise indices, liquidity/funding measures) should attenuate the market slope, especially on shock days. This is the most plausible baseline. The pitfall is to call $\beta$ “market impact’’ rather than “$Z$ exposure.’’

\item \textbf{Chain via the market (instrumental $Z$): $Z_t \to X_{t+\Delta} \to Y_{t+2\Delta}$.}  
An index-level actuator (e.g., binding funding/VAR/inventory constraints) transmits to constituents with a lag. If $Z$ does not directly move $Y$ (a strong exclusion), then $\beta^{\text{OLS}}_{2,1}$ identifies the direct causal $X\!\to\!Y$ effect. Empirically episodic (stress/liquidity regimes); claiming this graph requires lead–lag evidence and a concrete mechanism, not just a large contemporaneous $\beta$.

\item \textbf{Chain via the asset (constituent drives the index): $Z_t \to Y_{t+\Delta} \to X_{t+2\Delta}$.}  
Mega-cap news moves $Y$, then (via weights/flows) the market next interval. A regression $Y_{t+\Delta}\sim X_t$ is backward-looking; the causal direction is $Y\!\to\!X$. Using a leave-one-out index $X^{-i}$ should reduce the apparent “market’’ effect. This is common when index weights are concentrated.

\item \textbf{Collider at $Z$ (joint response): $X_t \to Z_{t+\Delta} \leftarrow Y_t$.}  
Here $Z$ is a response variable (e.g., realized volatility, order-book tightness, funding pressure) that reacts to both $X$ and $Y$. The regression $Y\sim X$ is unbiased, but conditioning on $Z$ creates spurious association between $X$ and $Y$ \parencite[collider bias]{pearl2016primer}. The practical pitfall is the “add more controls’’ reflex (e.g., controlling for VIX/liquidity) that opens this path.

\item \textbf{Market causes asset with other shocks (collider at $Y$): $X_t \to Y_{t+\Delta} \leftarrow Z_t$.}  
Other drivers of $Y$ act alongside a lagged $X$. If $X_t \perp (Z_t,U_{Y,t+\Delta})$ and $X$ is defined admissibly (lagged or leave-one-out), then $\beta^{\text{OLS}}_{1,0}$ can admit a causal reading as the direct $X\!\to\!Y$ effect (back-door blocked). Theoretically admissible, but empirically fragile unless a specific transmission mechanism is articulated.

\item \textbf{Asset causes market with other shocks (collider at $X$): $Y_t \to X_{t+\Delta} \leftarrow Z_t$.}  
Constituent moves---especially for large weights---propagate into the index next period, while broad forces also move $X$. Then $Y_t\sim X_{t+\Delta}$ is a backward-looking association; the causal direction is $Y\!\to\!X$. Expect stronger effects for mega-caps and attenuation when switching to $X^{-i}$.

\item \textbf{Latent confounding (unobserved $Z$): $Z_t \to X_{t+\Delta}, Z_t \to Y_{t+\Delta}$.}  
Persistent latent drivers (risk appetite, discount-rate drifts, sectoral shocks) induce $X$–$Y$ associations. OLS loads the shadow of $Z$; without good proxies, $\beta$ overstates “market causality.’’ This is common when the shock set is incomplete.

\end{enumerate}

\paragraph{Practical diagnostics before hedging.}
Two families of checks help separate these cases.  
First, \emph{timing}: test $X_{t-\Delta}\!\to\!Y_t$ versus $Y_{t-\Delta}\!\to\!X_t$ (Granger-style), controlling for $Z$ proxies. Chains via $X$ ((b)/(e)) should show $X$ leading $Y$; chains via $Y$ ((c)/(f)) should show the reverse.  
Second, \emph{measurement}: compare $\beta$ using $X_t$ versus $X_t^{-i}$ (leave-one-out); a large drop points to (c)/(f).  
Under forks ((a)/(g)), adding $Z$ proxies should materially attenuate the market slope on shock days; if it does not, either the proxies are weak or the DAG is mis-specified.  
Finally, avoid conditioning on response-type variables (colliders) like realized volatility or liquidity when estimating $Y\sim X$; doing so can manufacture dependence \parencite{pearl2016primer,de2023causal}.


\section{Numerical example (fork vs.\ chain)}\label{subsec:fork}

To make the discussion concrete, we compare two simple Gaussian structural causal models (SEM) that represent admissible stories for the CAPM slope:  
(i) a \emph{fork}, in which a common driver moves both the market and the asset, and  
(ii) a \emph{chain}, in which a market move causally propagates to the asset with correct time ordering.  

The fork illustrates classic errors-in-variables attenuation and shows that after “beta-neutralization’’ the residual still loads on the common driver.  
The chain, by contrast, yields a path coefficient exactly equal to OLS and a residual with no remaining driver load.  
These graphical claims follow directly from $d$-separation and back-door logic \parencite{pearl2009causality,pearl2016primer}.

Let $U_Z \sim \mathcal N(0,\sigma_Z^2)$, $U_X \sim \mathcal N(0,\sigma_X^2)$, and $U_Y \sim \mathcal N(0,\sigma_Y^2)$ be mutually independent exogenous shocks attached to $(Z,X,Y)$.

\subsection{Fork (common driver)}\label{subsec:fork-fork}

Consider the DAG $Z\to X$ and $Z\to Y$ (Section~\ref{sec:possible-dags}(a)).  
A compatible SCM is
\begin{equation}
Z := f_Z(U_Z),\quad
X := f_X(Z,U_X),\quad
Y := f_Y(Z,U_Y),
\end{equation}
and under linearity,
\begin{equation}
X \;=\; a Z + U_X,\qquad 
Y \;=\; b Z + U_Y,\qquad a\neq 0.
\end{equation}

Regressing $Y$ on $X$ yields
\begin{equation}
\beta_{\text{fork}}
= \frac{\Cov(Y,X)}{\Var(X)}
= \frac{ab\,\sigma_Z^2}{a^2\sigma_Z^2 + \sigma_X^2}
= \underbrace{\tfrac{b}{a}}_{\text{signal ratio}}
\times
\underbrace{\tfrac{a^2\sigma_Z^2}{a^2\sigma_Z^2 + \sigma_X^2}}_{\lambda\in(0,1)}.
\end{equation}
Thus $\beta_{\text{fork}}\to b/a$ (the signal ratio) only when $X$ is a noiseless proxy for $Z$ ($\sigma_X^2\to 0$).  
With measurement noise, $\beta$ is attenuated by the usual errors-in-variables factor $\lambda$.

After “beta-neutralization’’ the residual still loads on the driver: \begin{equation} \label{eq:resid-fork} \widetilde Y := Y - \beta_{\text{fork}} X = \bigl(b - a\,\beta_{\text{fork}}\bigr) Z + \bigl(U_Y - \beta_{\text{fork}}U_X\bigr), \qquad b - a\,\beta_{\text{fork}} = b\,\frac{\sigma_X^2}{a^2\sigma_Z^2 + \sigma_X^2}. \end{equation} Hence $\Cov(\widetilde Y,Z)=\bigl(b-a\,\beta_{\text{fork}}\bigr)\sigma_Z^2\neq 0$ whenever $\sigma_X^2>0$ and $b\neq 0$.
In words: hedging against the market index does not remove the underlying driver.

\subsection{Chain (market\,\texorpdfstring{$\to$}\,asset)}\label{subsec:fork-chain}

Now consider $Z\to X \to Y$ (Section~\ref{sec:possible-dags}(b)) with correct time ordering (or, equivalently, a leave-one-out market to avoid the aggregator issue):
\begin{equation}
X_t \;=\; a Z_{t-\Delta} + U_{X,t},\qquad 
Y_t \;=\; c\,X_{t-\Delta} + U_{Y,t}.
\end{equation}
With $(U_{X},U_{Y},Z)$ independent across $t$, the regression
\begin{equation}
\beta_{\text{chain}} = \frac{\Cov(Y_t,X_{t-\Delta})}{\Var(X_{t-\Delta})} = \frac{\Cov(c\,X_{t-\Delta}+U_{Y,t},X_{t-\Delta})}{\Var(X_{t-\Delta})} = c
\end{equation}
recovers the true causal effect exactly.  

Here the beta-hedged residual eliminates the driver ($Z$):
\begin{equation}
\widetilde Y_t := Y_t - \beta_{\text{chain}}X_{t-\Delta} = U_{Y,t},
\end{equation}
which is orthogonal to $Z_t$ for all $t$.  
Unlike the fork, hedging here removes the mechanism cleanly.

\subsection{Numerical simulation}\label{subsec:sim}

We simulate both SEMs to illustrate these contrasts.  
In the fork:
\[
X_t \;=\; a\,Z_{t-\Delta} + U_{X,t},\qquad
Y_t \;=\; b\,Z_{t-\Delta} + U_{Y,t},
\]
so $X$ is a noisy proxy for $Z$.  
In the chain:
\[
X_t \;=\; a\,Z_{t-\Delta} + U_{X,t},\qquad
Y_t \;=\; c\,X_{t-\Delta} + U_{Y,t}.
\]

We draw $n=100{,}000$ observations with $a=b=c=1$, $\sigma_Z=\sigma_Y=1$, and vary $\sigma_X\in[0,5]$.  
For each $\sigma_X$ we estimate the OLS slope and---in the fork---the post-hedge loading on $Z$.  

\begin{figure}[t]
\centering

\hspace*{\fill}%
\begin{subfigure}[t]{0.32\textwidth}
  \centering
  \includegraphics[width=\linewidth]{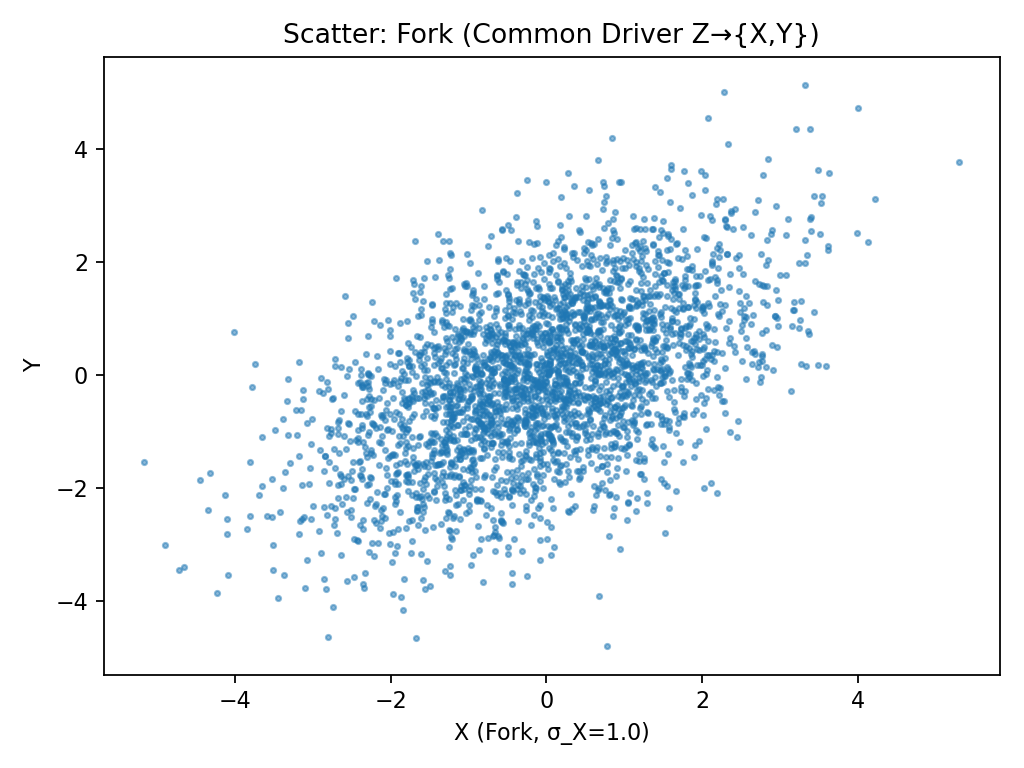}\vspace{-0.35em}
  \caption{Fork: contemporaneous co-movement $Y$ vs.\ $X$.}
\end{subfigure}\hfill
\begin{subfigure}[t]{0.32\textwidth}
  \centering
  \includegraphics[width=\linewidth]{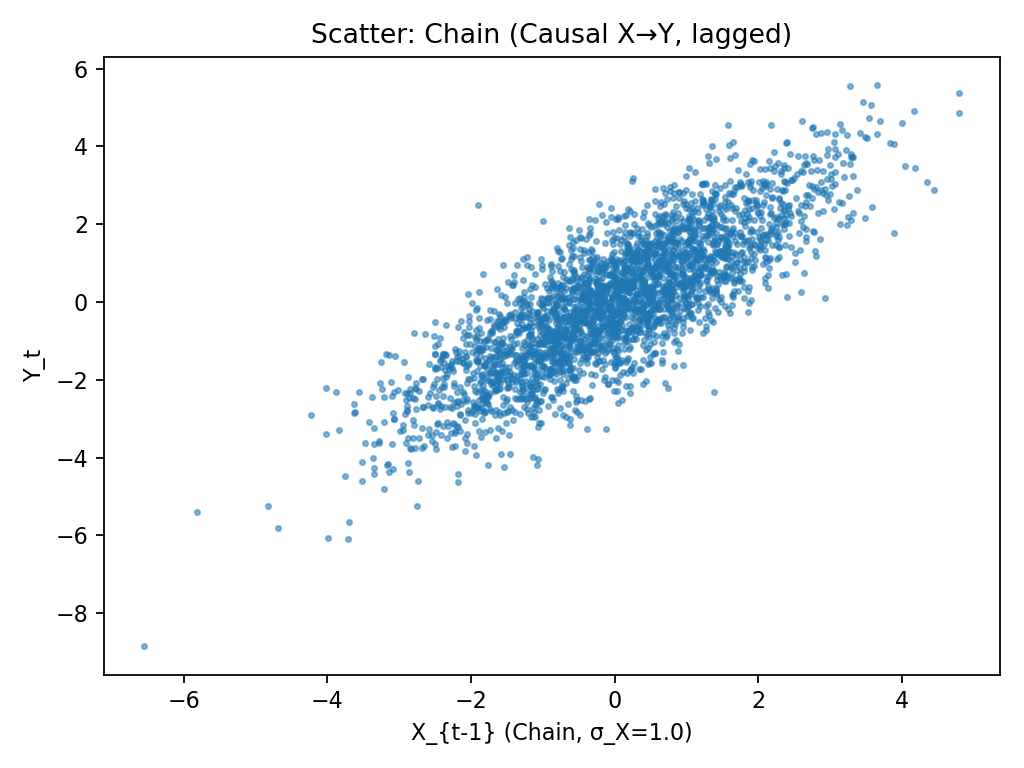}\vspace{-0.35em}
  \caption{Chain (lagged): $Y_t$ vs.\ $X_{t-\Delta}$.}
\end{subfigure}%
\hspace*{\fill}

\vspace{0.8em}

\begin{subfigure}[t]{0.32\textwidth}
  \centering
  \includegraphics[width=\linewidth]{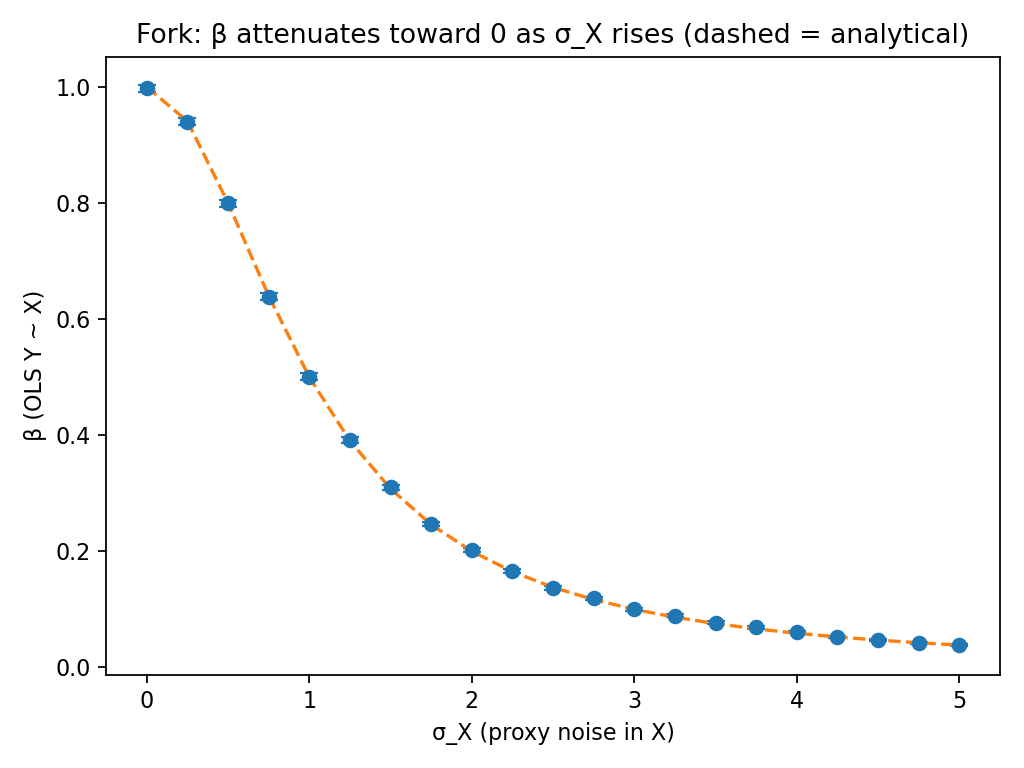}\vspace{-0.35em}
  \caption{Fork: $\beta_{\text{OLS}}$ attenuates toward $0$ as $\sigma_X$ increases.}
\end{subfigure}\hfill
\begin{subfigure}[t]{0.32\textwidth}
  \centering
  \includegraphics[width=\linewidth]{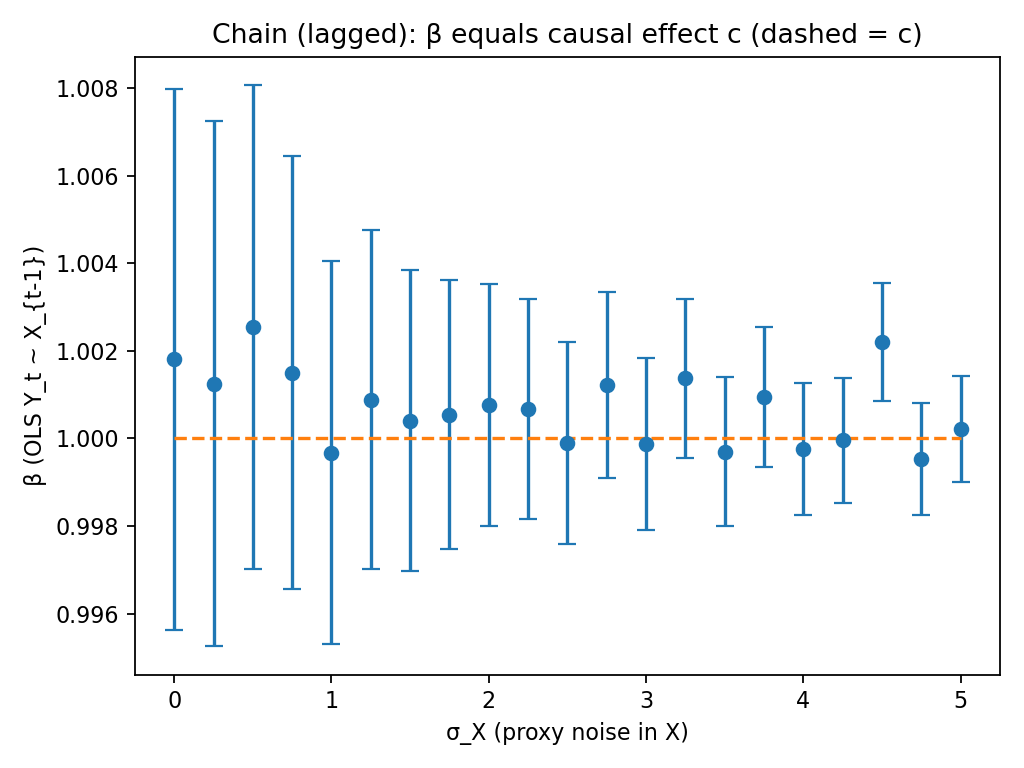}\vspace{-0.35em}
  \caption{Chain (lagged): $\beta_{\text{OLS}}$ equals $c$ for all $\sigma_X$.}
\end{subfigure}\hfill
\begin{subfigure}[t]{0.32\textwidth}
  \centering
  \includegraphics[width=\linewidth]{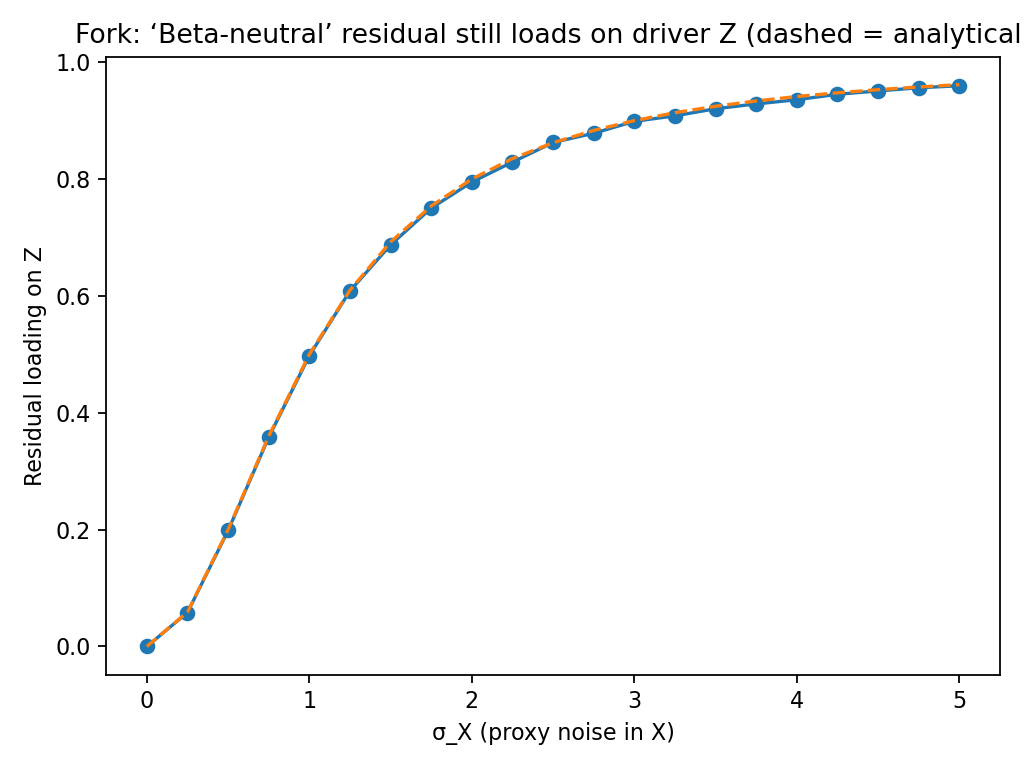}\vspace{-0.35em}
  \caption{Fork: “beta-neutral’’ residual’s loading on $Z$ grows with $\sigma_X$.}
\end{subfigure}

\caption{\textbf{Fork vs.\ chain (simulation).}
Top: scatterplots illustrate contemporaneous co-movement in a fork and a lagged causal mechanism in a chain.  
Bottom: in a fork, $\beta$ is an attenuated proxy and hedging does not remove the driver; in a chain, $\beta$ equals the causal effect and hedging removes the mechanism.  
Monte Carlo estimates (points) track the analytical values (dashed lines).}
\label{fig:numerics-fork-chain}
\end{figure}

The results (Fig.~\ref{fig:numerics-fork-chain}) mirror the algebra.  
In the fork, $\beta$ attenuates smoothly with $\sigma_X$ and the residual’s $Z$-loading grows toward $b$ as $\sigma_X$ rises.  
In the chain, $\beta$ recovers $c$ for all $\sigma_X$ and the hedge removes the driver completely.  

\section{Empirical checks of the minimal graph}\label{sec:empirical}

\noindent
We ask whether observed returns behave as they must if the contemporaneous CAPM slope is a proxy for shared drivers (a fork $Z\!\to\!\{R_m,R_i\}$), and whether any genuine market$\to$stock mechanism appears only when time ordering is respected. These are falsifiable consistency checks, not identification exercises.

\paragraph{Data and setup.}
We use U.S.\ equities and macro data from public sources (ALFRED; Yahoo Finance), 2015–2025. The market proxy $R_m$ is the SPY log return. The cross-section comprises $\sim\!100$ large-cap U.S.\ tickers, balanced across the 11 SPDR sectors. As observable proxies for the driver $Z$, we include the daily change in VIX ($\Delta$VIX), the change in the 10y Treasury yield ($\Delta$DGS10), and the DXY log return; we also add the corresponding SPDR sector ETF return as a coarse sector control.%
\footnote{We do not build leave–one–out indices; the objective here is to assess contemporaneous $\beta$ strictly as a proxy.}
Consumer Price Index (CPI) release dates provide exogenous macro “shock” days. 

\subsection*{Pillar 1: contemporaneous $\beta$ behaves like a proxy}

If $\beta$ reflects exposure to a common driver $Z$, three patterns should hold:  
(i) on shock days, adding observable $Z$ proxies should strongly attenuate the $R_m$ slope;  
(ii) the slope should vary systematically across environments;  
(iii) after beta–hedging, residuals should still load on macro/sector shocks, since the hedge only removes a projection on another child of $Z$, not $Z$ itself.

\begin{figure}[t]
\centering

\begin{subfigure}[t]{0.32\textwidth}
  \centering
  \includegraphics[width=\linewidth]{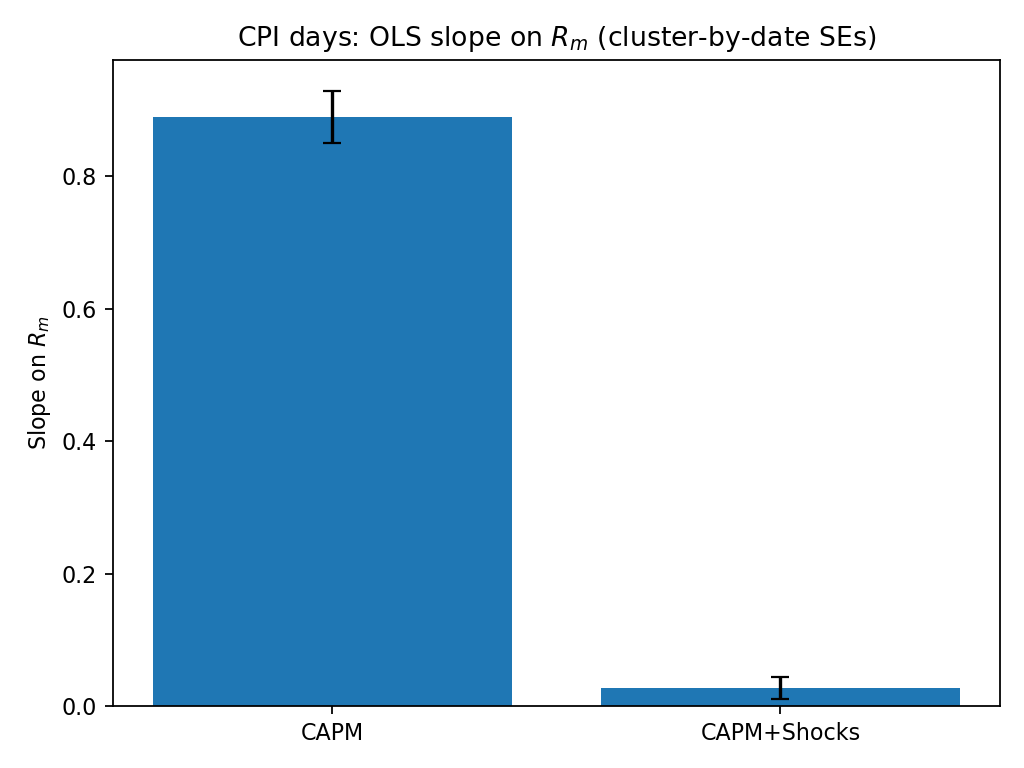}\vspace{-0.35em}
  \caption{CPI days: pooled slope on $R_m$ with/without shock controls.}
\end{subfigure}\hfill
\begin{subfigure}[t]{0.32\textwidth}
  \centering
  \includegraphics[width=\linewidth]{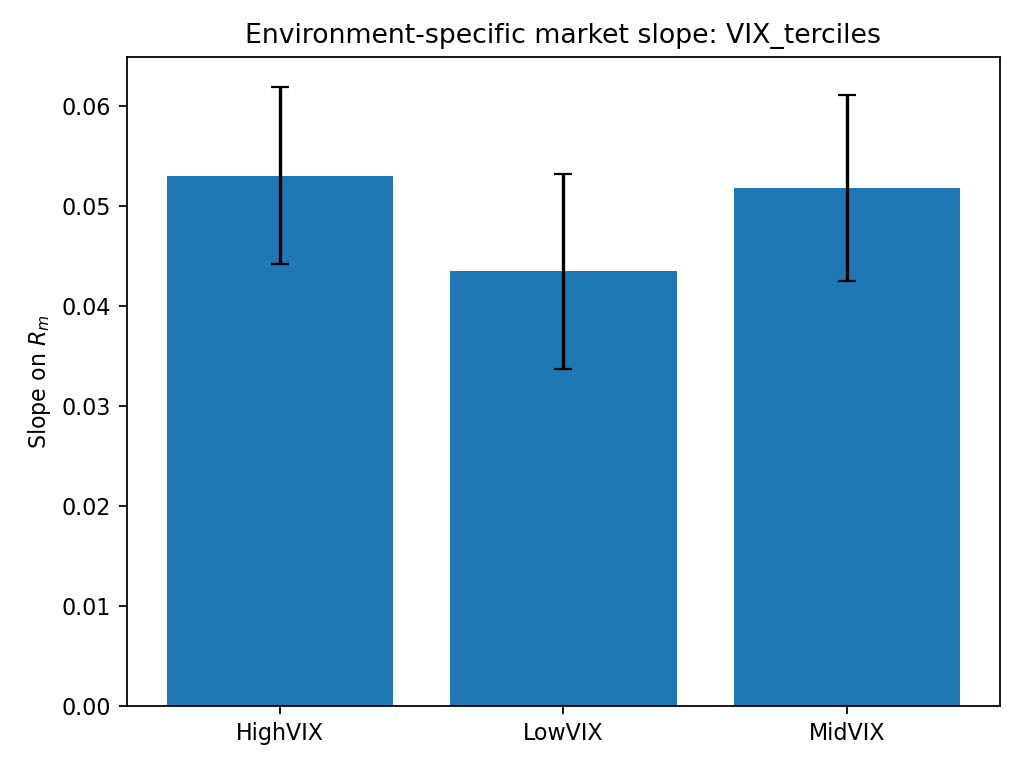}\vspace{-0.35em}
  \caption{$\beta_e$ by VIX terciles.}
\end{subfigure}\hfill
\begin{subfigure}[t]{0.32\textwidth}
  \centering
  \includegraphics[width=\linewidth]{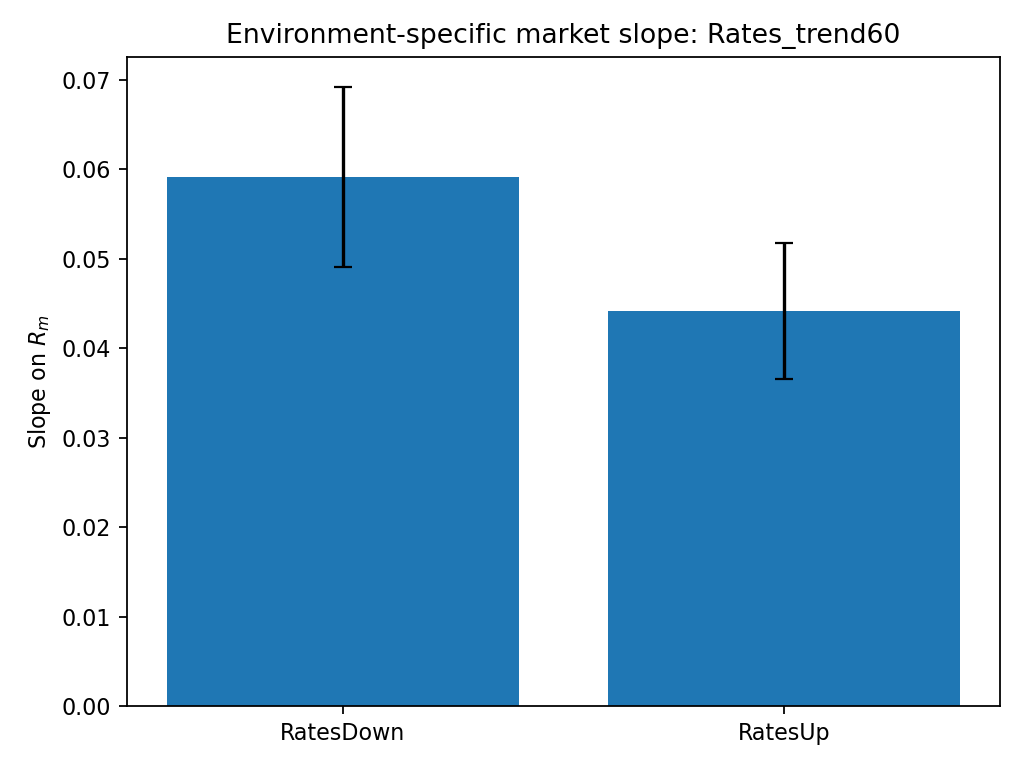}\vspace{-0.35em}
  \caption{$\beta_e$ by 60-day rates trend.}
\end{subfigure}

\vspace{0.8em}

\begin{subfigure}[t]{0.32\textwidth}
  \centering
  \includegraphics[width=\linewidth]{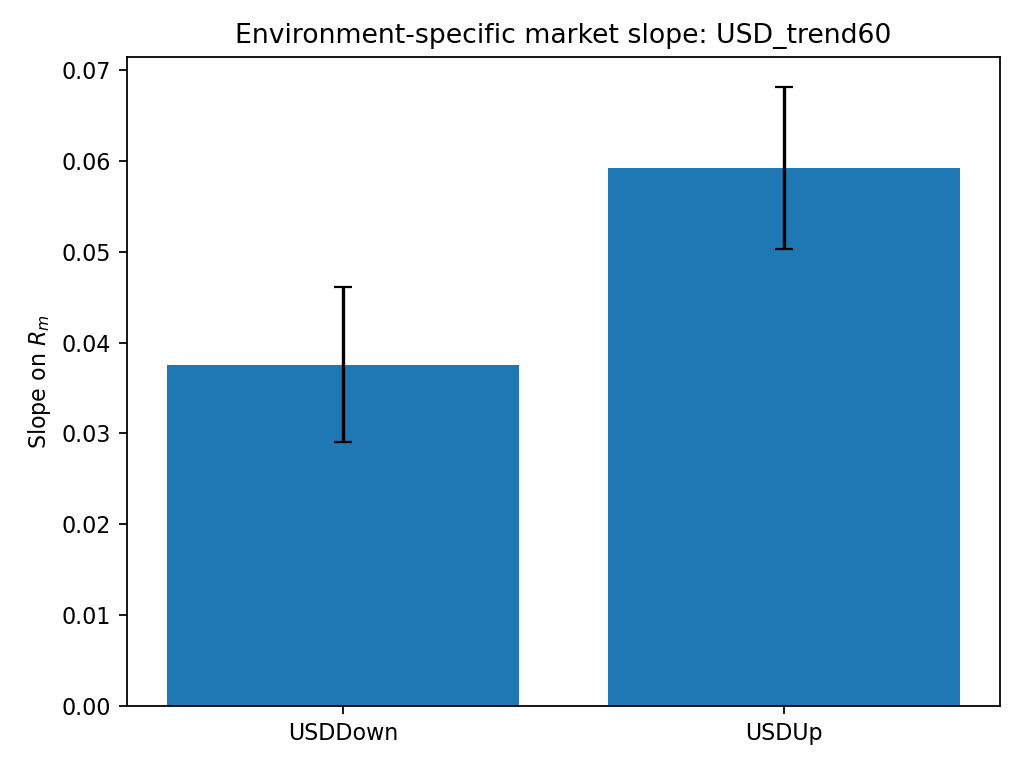}\vspace{-0.35em}
  \caption{$\beta_e$ by 60-day USD trend.}
\end{subfigure}\hfill
\begin{subfigure}[t]{0.32\textwidth}
  \centering
  \includegraphics[width=\linewidth]{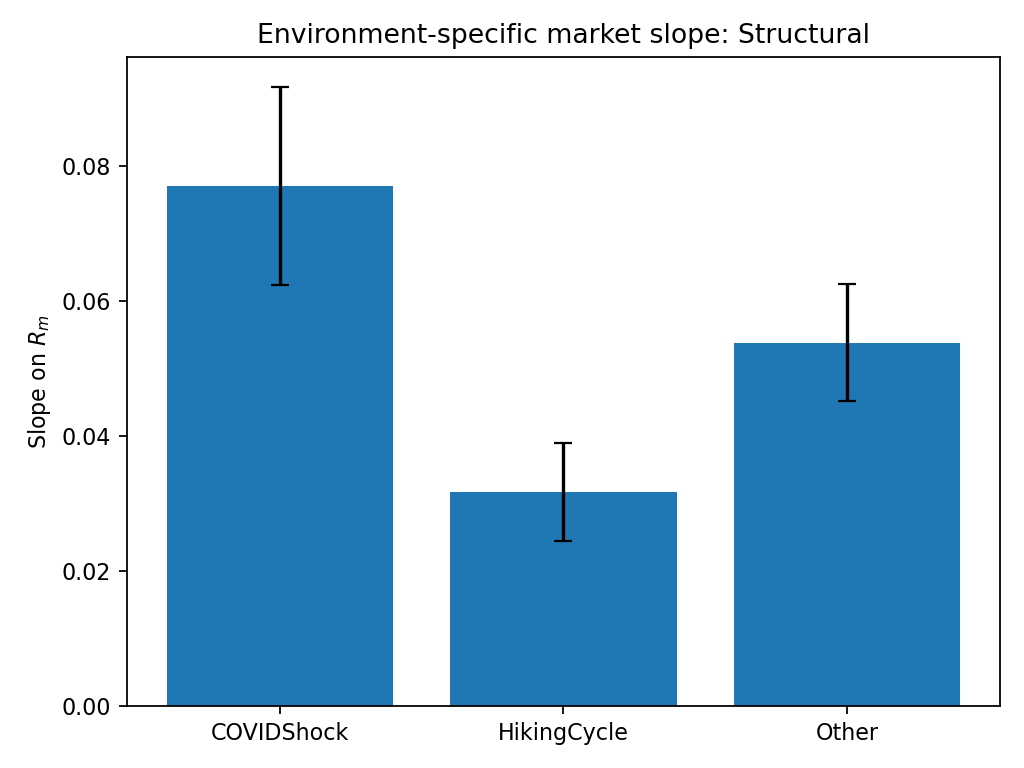}\vspace{-0.35em}
  \caption{$\beta_e$ by structural episode.}
\end{subfigure}\hfill
\begin{subfigure}[t]{0.32\textwidth}
  \centering
  \includegraphics[width=\linewidth]{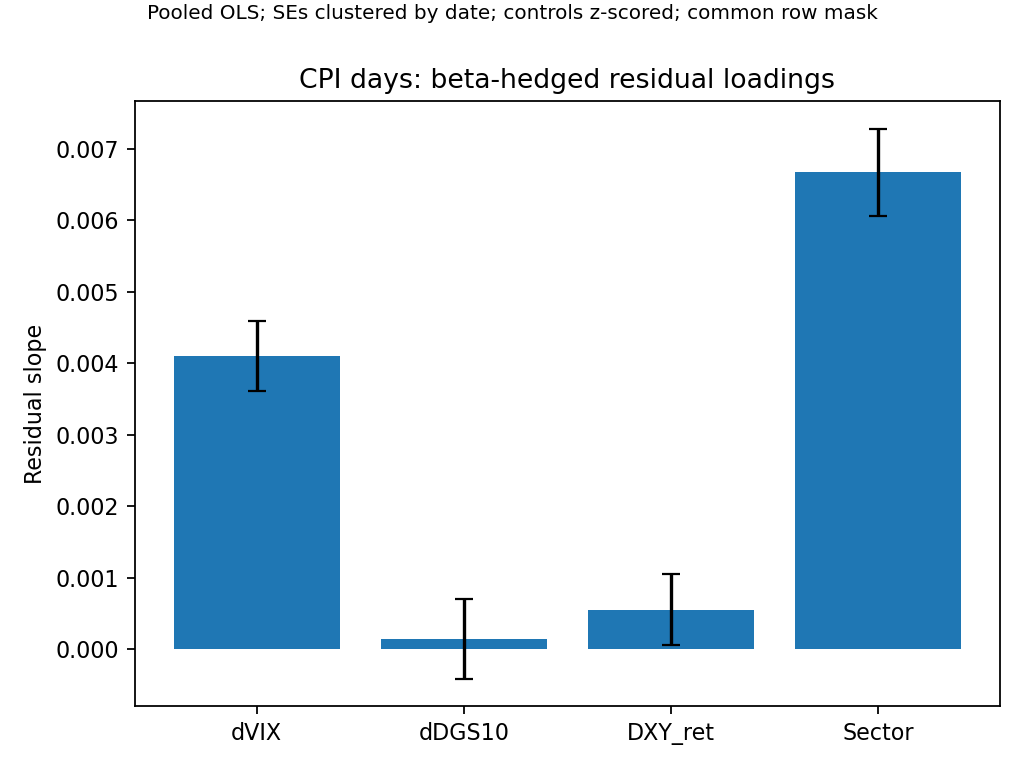}\vspace{-0.35em}
  \caption{CPI days: beta-hedged residual loadings on standardized shocks.}
\end{subfigure}

\caption{\textbf{Pillar 1---contemporaneous CAPM behaves like a proxy.}
\emph{(a):} adding observable shocks (VIX, rates, USD, sector) collapses the $R_m$ slope on CPI days.  
\emph{(b–e):} the market slope $\beta_e$ varies materially across volatility, rates, USD, and structural regimes.  
\emph{(f):} after hedging, residuals still load on macro/sector shocks. This is inconsistent with a stable same‐day $R_m\!\to\!R_i$ mechanism and consistent with a fork.}
\label{fig:pillar1}
\end{figure}

\medskip\noindent
\textit{(i) CPI days---control robustness.}  
On CPI releases we pool the cross‐section and compare $R_{i,t}$ on $R_{m,t}$ alone versus $R_{m,t}$ plus $\{\Delta\mathrm{VIX},\Delta\mathrm{DGS10},\mathrm{DXY}^{\mathrm{ret}},\mathrm{Sector}^{\mathrm{ret}}\}$.  
Figure~\ref{fig:pillar1}(a) shows order–of–magnitude attenuation in the $R_m$ slope (from $\sim 0.9$ to $\sim 0.03$). This is difficult to reconcile with a stable same–day causal mechanism and aligns with $R_m$ acting as a noisy proxy for $Z$.

Because sector ETFs are strong proxies for single–stock variation, we re–estimate the specification without sector ETFs (not shown). Excluding sector factors leaves the $R_m$ slope positive—though still materially attenuated—after controlling for macro shocks.
This distinction highlights that attenuation can arise from (i) causal structure (a fork) and (ii) proxy competition among regressors.
We therefore treat sector factors as an informative but optional diagnostic: if the market slope collapses only when sectors are included, that pattern points to proxy dominance rather than an intrinsic market$\to$asset causal link.

\medskip\noindent
\textit{(ii) Environment dependence.}  
We estimate pooled interaction regressions,
\[
R_{i,t}=\alpha + \beta\,R_{m,t} + \sum_{e\neq e_0}\!\big(\gamma_e\,\mathbf{1}\{E_t{=}e\} + \delta_e\,R_{m,t}\,\mathbf{1}\{E_t{=}e\}\big) + \Gamma^\top Z^{\text{ext}}_t + u_{i,t},
\]
and recover environment–specific slopes $\beta_e=\beta+\delta_e$.  
Figures~\ref{fig:pillar1}(b–e) show systematic differences: larger slopes in \emph{COVIDShock}, \emph{RatesDown}, and during USD uptrends (\emph{USDUp}); smaller slopes in the hiking cycle, \emph{USDDown}, and during rate uptrends. Such state dependence is characteristic of proxy behavior, not an invariant mechanism.

When re-estimated without sector ETFs (not shown), the environment patterns persist but the absolute market slopes are somewhat larger, indicating that sector factors absorb part of the variation yet do not create the state dependence itself.

\medskip\noindent
\textit{(iii) Post–hedge residual loadings.}  
For each CPI date we form residuals using a pre‐event rolling CAPM and regress them on the standardized shock controls.  
Figure~\ref{fig:pillar1}(f) shows clear loadings on $\Delta$VIX and sector returns, with heterogeneous patterns across CPI vintages. A genuine same–day $R_m\!\to\!R_i$ mechanism would not morph in this way; the evidence is consistent with a fork $Z\!\to\!\{R_m,R_i\}$.

Excluding sector ETFs (not shown) leaves the $\Delta$VIX loading prominent, confirming that residual co-movement with macro shocks remains even without sector competition.

\subsection*{Pillar 2: any market$\to$stock mechanism must be lagged}

\begin{figure}[t]
  \centering
  \includegraphics[width=0.56\textwidth]{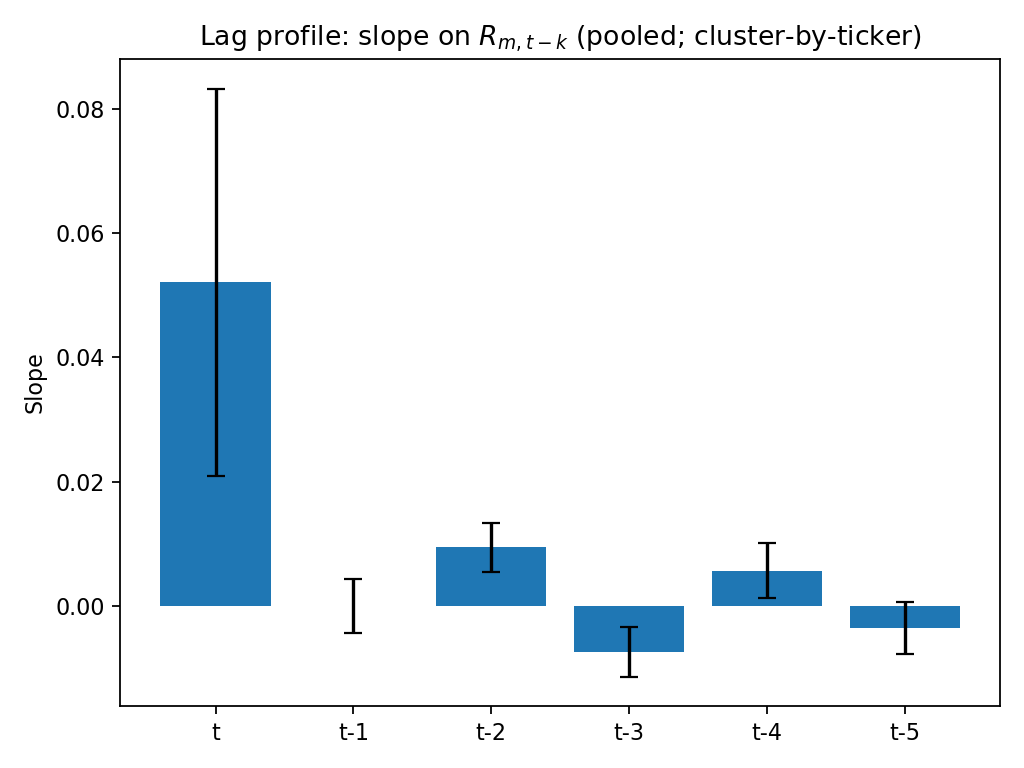}
  \caption{\textbf{Pillar 2---lag profile of the market slope.}
  Pooled slopes on $R_{m,t-k}$ for $k=0,\dots,5$. The contemporaneous association ($k=0$) is large and non-causal; lagged slopes are small and unstable.}
  \label{fig:pillar2}
\end{figure}

We next estimate a pooled lag profile with the same contemporaneous controls,
\[
R_{i,t}=\alpha+\sum_{k=0}^{5} c_k\,R_{m,t-k} + \Gamma^\top Z^{\text{ext}}_{t} + e_{i,t}.
\]
Figure~\ref{fig:pillar2} shows the pattern: $c_0$ is large, reflecting the contemporaneous proxy relation. At $k\ge 1$, slopes are small, alternate in sign, and most confidence intervals straddle zero. If anything, there is a weak, sample-specific blip at $k{=}2$ that does not persist. This is what one would expect if flow/inventory channels exist but are economically second order relative to common drivers at this resolution.

\medskip\noindent
Across both pillars the data look like a fork. On shock days the market slope collapses once $Z$ proxies are added, slopes vary by environment, and residuals after beta-hedging still load on shocks. Any genuine market$\to$stock actuation, if present, appears only at a lag and at much smaller scale. This is exactly the empirical signature implied by the minimal DAG where $Z$ drives both $R_m$ and $R_i$.


\section{Discussion \& Conclusions: What a Causal Lens Changes}\label{sec:conclusion}

\noindent
This paper asked a simple question: what does the CAPM regression slope represent once realized returns are viewed through an acyclic structural causal model (SCM)? The answer is clear: a same-period CAPM cannot be a structural model, because the market index is itself defined as a contemporaneous aggregate of its constituents. Without respecting time ordering, the regression slope is not a causal effect but a projection whose slope shadows the drivers that move both the index and the stock. The issue is methodological, not semantic: causal claims require arrows justified by timing, mechanism, and elimination properties, not just high $R^2$ or familiar factor menus \parencite{pearl2016primer,de2023causal}.

\paragraph{What the aggregator identity rules out.}
With $X\equiv R_m$ constructed as $X_t=\sum_j w_{j,t-1}Y_{j,t}$, positing a same-time arrow $X_t\!\to\!Y_{i,t}$ creates a cycle unless either $w_{i,t-1}=0$ (leave-one-out) or the effect is zero (\S\ref{subsec:agg-contr}). This is not a technicality but a structural incompatibility: “beta’’ cannot be a same-day physical actuator of $Y$. In a causal model, edges must earn their keep—they must be temporally admissible, acyclic at the sampling index, supported by a concrete transmission mechanism, and satisfy the causal elimination property (\S\ref{subsec:necessary1}).

\begin{figure}[t]
\centering
\tikzset{dot/.style={circle,fill=black,inner sep=1.2pt}}
\begin{tikzpicture}[>=stealth, node distance=20mm]
\node[dot,label=left:{$Z_t$}] (Z) {};
\node[dot,label=right:{$R_{m,t+\Delta}$}, right=of Z, yshift=6mm] (X) {};
\node[dot,label=right:{$R_{i,t+\Delta}$}, right=of Z, yshift=-6mm] (Y) {};
\draw[->] (Z) -- (X);
\draw[->] (Z) -- (Y);
\end{tikzpicture}

\caption{\textbf{CAPM through a causal lens.}  
Same-period CAPM is incoherent as a causal SCM: it violates temporal priority, acyclicity, mechanism, and causal elimination.  
The plausible baseline is a fork $Z_t\!\to\!\{R_{m,t+\Delta},R_{i,t+\Delta}\}$, making the $R_m$–$R_i$ relation \emph{associative}, not causal.  
Here $\beta$ is an attenuated proxy for $Z$, so hedging on $R_m$ leaves $Z$ exposure.}
\label{fig:final-fork}
\end{figure}
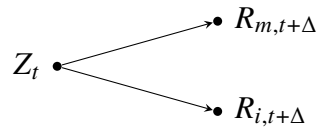

\paragraph{What remains plausible.}
Once those necessities are enforced, the empirically dominant baseline is a fork  (Fig.~\ref{fig:final-fork}): a common driver $Z$ (macro news, discount-rate moves, liquidity or risk appetite) pushes both $R_m$ and $R_i$. In that DAG, the CAPM slope is a directional projection on another child of $Z$, attenuated by how noisily $R_m$ measures the shock. Beta-neutral residuals still load on $Z$. Our numerical examples make this transparent: in a fork, $\beta$ attenuates and residuals retain $Z$ exposure; in a lagged chain, OLS recovers the true causal effect and the hedge removes the mechanism (\S\ref{subsec:fork}; Fig.~\ref{fig:numerics-fork-chain}).

\paragraph{What the data show.}
The daily evidence fits this fork pattern. On CPI days, adding observable shocks collapses the slope on $R_m$. Across environments (volatility terciles, rate/FX trends, structural regimes), $\beta$ varies systematically, inconsistent with an invariant mechanism. After hedging, residuals still load on shocks, with heterogeneous contributions from VIX, sector, rates, and USD. A true same-day market$\to$stock mechanism would not survive hedging in this way. When we look for genuine actuation, it appears—if at all—only at lags, and even then with small, unstable coefficients. This leaves room for episodic, stress-induced channels (inventory, VAR, margin constraints) but little support for a standing same-day arrow.

\paragraph{What this changes for practice.}
First, treat CAPM slopes as \textit{descriptive} unless the model is explicitly causal: that requires lagged or leave-one-out indices and a concrete actuation story.  
Second, reframe hedging as path surgery rather than menu subtraction: specify the DAG, name the horizons, block back-doors, and hedge the mechanism relevant to your query.  
Third, reserve “alpha’’ for what remains invariant once the causal paths you have declared are blocked and falsified across environments. Only then does the residual merit that label.

\paragraph{Scope and limits.}
Our scope is deliberately narrow: period returns, a minimal three-node core, and linear SEM illustrations. Microstructure and institutional frictions can create short-lived arrows; those require finer timing, leave-one-out constructions, and explicit actuation (who pushes, through what instrument, and when). Our $Z$ proxies are noisy; attenuation on shock days is precisely the fork prediction. None of our checks “identify the DAG from prices’’; they filter admissible graphs and expose incoherent ones—the role of causal modeling before causal estimation.

\medskip\noindent
The message is simple. Treating CAPM as causal at the same sampling index is a category error. The right unit of analysis is the mechanism. Specify the SCM, earn the arrows with necessities, and reserve “alpha’’ for what remains invariant once you excise the causal paths you have declared. Everything else is association with a story.

\section*{Acknowledgments}
We thank Alex Dannenberg for thoughtful comments on horizon choice and on attenuated market proxies in concentrated indices.

\section*{Disclosure}
The views expressed are solely the author’s and do not necessarily reflect those of any current or former employer or affiliated institution. This paper was prepared in a personal capacity, without the use of employer resources or confidential information. No investment advice is offered. Any errors are the author’s alone.


\printbibliography

\end{document}